\RequirePackage{ifpdf}
\documentclass{PoS}

\title{Closure testing NNPDF3.0 with LHC observables}

\ShortTitle{Closure testing NNPDF3.0 with LHC observables}

\author{\speaker{Christopher Deans}%
        \thanks{On behalf of the NNPDF collaboration: R.~D. Ball, V.~Bertone, S.~Carrazza, C.~S. Deans, L.~Del~Debbio, S.~Forte, P.~Groth~Merrild, A.~Guffanti, N.~P. Hartland, Z.~Kassabov, J.~I. Latorre, J.~Rojo, L.~Rottoli, M.~Ubiali}\\
       University of Edinburgh\\
       E-mail: \email{c.s.deans@sms.ed.ac.uk}}

\abstract{A thorough understanding of PDFs and their uncertainties is important for the LHC and for future collider experiments. The recently released NNPDF3.0 set was presented alongside results from closure tests, where PDF fits were performed on pseudo-data generated from a chosen input PDF set. The results there demonstrate the validity of the NNPDF methodology and also provide some information about different contributions to the PDF uncertainties. Here I present a number of additional closure test results, specifically an investigation into the effect of using cross-validation in the fits, and an assessment of the successful reproduction of LHC cross-sections in closure tests. The results are consistent with those previously shown in the NNPDF3.0 paper.}

\FullConference{XXIII International Workshop on Deep-Inelastic Scattering\\
		 27 April - May 1 2015\\
		 Dallas, Texas}

\begin{document}

\section{Introduction}

As we enter the era of the 13 TeV LHC, it is increasingly important that parton distribution functions and their uncertainties are well understood. 
Already for a number of key LHC measurements PDF uncertainties are of equivalent size to the experimental uncertainties~\cite{ForteWatt}, and this situation will get worse as more data is collected, and as the size of other theory uncertainties (e.g.~from the scale) are reduced.
To face this challenge, new global determinations of PDFs have recently been performed: NNPDF3.0~\cite{NN30}, MMHT2014~\cite{MMHT}, and the upcoming CT14.
Due to improvements in the methodology and theory used by the various fitting groups, discrepancies between their results highlighted previous benchmarking exercises~\cite{pdfbench} have reduced, resulting in many cases in improved agreement between the new sets.
This better agreement has also led to the development of combined PDF sets~\cite{cmcpdf}.
However, it is still important to test statistically that the methodologies used to perform the PDF fits are valid, and produce a result which is unbiased.

As part of the NNPDF3.0 analysis, we performed a large number of closure tests on our fitting methodology.
As described later in this contribution, these tests involved fitting a set of pseudo-data generated from a chosen PDF set, with the aim of assessing whether fit successfully reproduced the supplied underlying law, up to statistical fluctuations.
Results of many of these closure tests were presented in the main NNPDF3.0 paper~\cite{NN30}; in this proceedings I will present a number of additional results, specifically of closure tests using reduced datasets, and for LHC observables.

\section{NNPDF3.0}

NNPDF3.0, our latest PDF determination, was released in October 2014~\cite{NN30} and features new data from the LHC and HERA, and a completely updated analysis code and methodology.
Amongst the new data included were HERA-II structure function data from both H1 and ZEUS, and HERA combined charm production cross-section data.
Adding to the LHC data already included in the NNPDF2.3 analysis~\cite{NNPDF2.3}, NNPDF3.0 introduced a large amount of the released LHC run-I data, including many new processes important to PDFs. 
We added the ATLAS 2.76 TeV inclusive jet data with systematics fully correlated to the previously included 7 TeV ATLAS set, which increases the impact of both sets in the fit.
Also new were the $W+c$ data from CMS, which provide important information on the strange PDFs, and data on the total top pair production cross-section from both ATLAS and CMS, making use of the recent result for the full NNLO calculation~\cite{rojo:top}.

For this new determination we also renovated our fitting methodology.
The NNPDF3.0 fits were performed using a completely new code, written in C++ and optimised for the computational intensive hadronic calculations required for our fits.
The genetic algorithm we use to perform the minimisation was updated, with a new mutation approach which makes use of the structure of the neural networks to improve both fitting speed and results.
We extended the number and kinematic range of positivity observables we used in the fits, in order to better constrain unphysical negativity in the PDFs.
As I will discuss further in Section~\ref{sec:cv}, NNPDF3.0 also features an improved form of cross-validation, which prevents over-learning in the neural networks but with a reduced chance of under-learning due to premature stopping of the fit.

The NNPDF3.0 sets are available on {\sc\small LHAPDF}~\cite{LHAPDF}, with determinations at LO, NLO and NNLO for multiple values of $\alpha_S$, and also for a number of reduced datasets.
More details about the new data included and the new methodological features are available in the paper~\cite{NN30}.

\section{Closure testing}

One of the central features of the NNPDF3.0, which was also central to the development of the new methodology, is the use of closure tests.
The basic idea of these tests is to perform a fit where we know the underlying `correct' answer, and so allowing us to directly evaluate how accurate our fits are.
This technique was used in NNPDF3.0 both to validate our results and to test improvements to the methodology.

In order to perform a closure test, first a set of pseudo-data is generated using a chosen input PDF set.
The pseudo-data we used was based on the NNPDF3.0 dataset, using the same covariance matrix but with central values based on the theory value from the PDF set, fluctuated according to the experimental uncertainties.
Different central values can be obtained by using different random seeds in this process, so the closure test is repeatable.
There are also a number of different options at this stage.
For instance, a variation on the standard closure tests can be done where the pseudo-data central values are generated without statistical fluctuations, i.e.~are set as the pure theory value from the input PDFs.
This provides an environment to test the neural network minimisation where over-learning is impossible, and features can be evaluate purely on the goodness-of-fit they obtain.
Once the pseudo-data is generated with the chosen settings, a PDF fit can be performed with it in exactly the same way as with the real experimental data.

Given how the pseudo-data is generated, it is automatically perfectly consistent with the theory used to generate it, and so the precise theoretical choices (perturbative order, quark mass thresholds, value of $\alpha_S$) do not affect the results as long as the same settings are used in both cases.
For the closure test presented here and in~\cite{NN30}, we use the default NNPDF3.0 NLO settings.
One related issue is the use of positivity constraints in the closure test fits.
If these are not satisfied by the input PDFs, including them in the fit could introduce some tension between them and the pseudo-data.
On this basis we do not include the positivity constraints where this is the case.

\section{Cross validation}
\label{sec:cv}

One key improvement in the NNPDF methodology introduced for NNPDF3.0 is improved cross-validation.
The idea of cross-validation is simple: in order to prevent the neural networks from over-learning, split the dataset into two halves, training the networks with one while monitoring the quality of the fit to the other, validation, set.
While the $\chi^2$ is improving to both the training and validation sets, this means that the neural network is correctly learning the underlying law, while when it improves for the training set and deteriorates for the validation set, the neural networks are overfitting.
In previous analyses, we have monitored the validation $\chi^2$ during the minimisation and stopped the fit when it started to increase.
However, the validation $\chi^2$ is subject to a substantial amount of noise, and this approach could lead to the fit being stopped too earlier, resulting in under-learning.
This was the case even when approaches to reduce the noise, for example smoothing the $\chi^2$ over several generations, were used.

The new approach does not attempt to stop the fit, and instead allows it to continue for a preset large number of generations.
The best generation, in terms of over- and under-learning, is then selected from all seen during the fit as the generation with the lowest validation $\chi^2$.
This technique therefore prevent overfitting by taking as the final result the set of networks which has the best fit to a set of unseen data, while avoiding the previously mentioned issues with the standard stopping approach.


\begin{figure}
\centering
\includegraphics[width=0.95\textwidth]{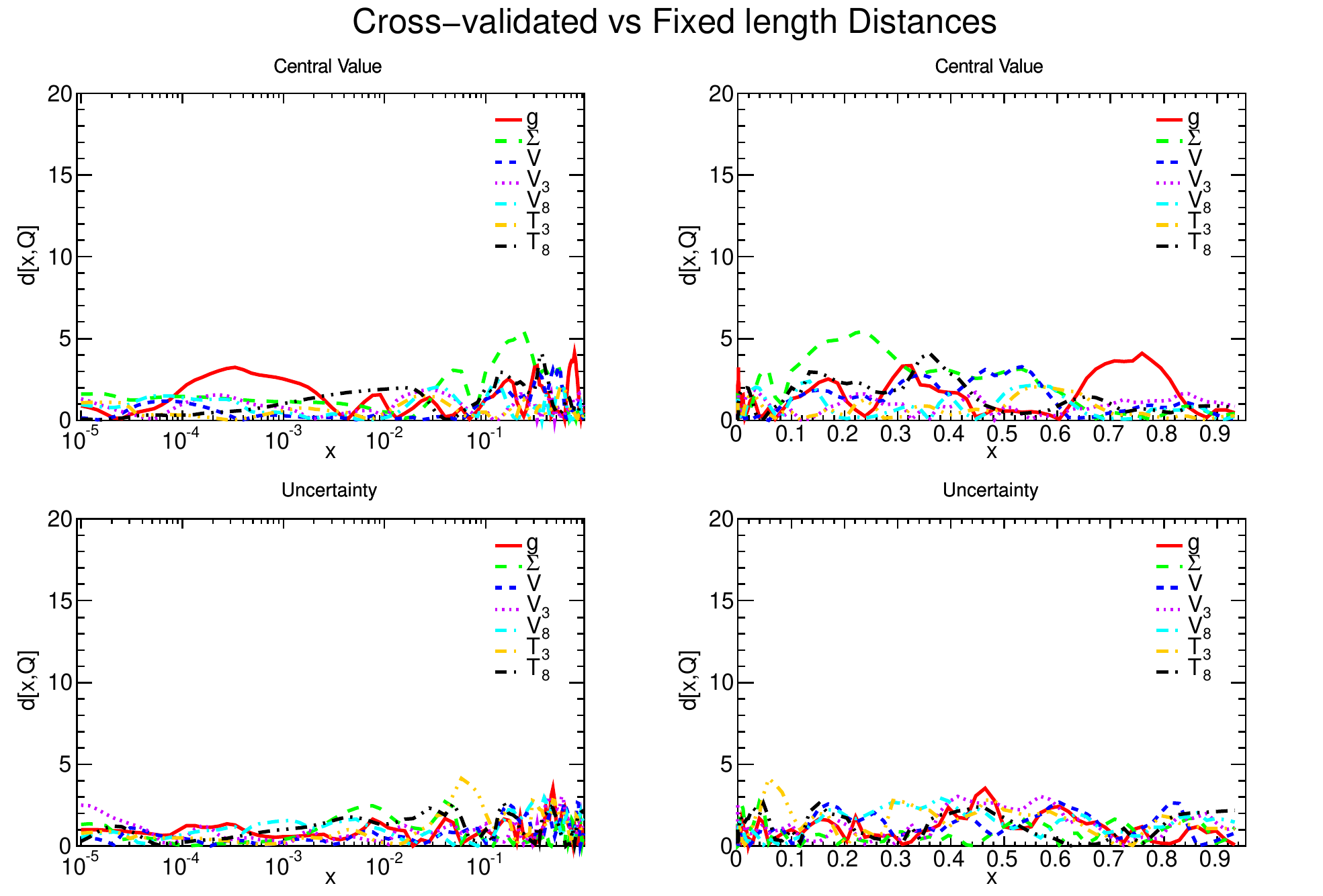}
\caption{\small
PDF distances between closure test fits with and without cross-validation.
The closure test fits were performed using pseudo-data based on MSTW2008.
}
\label{fig:diffdistance-cv-fl}
\end{figure} 

\begin{figure}[h]
\centering
\includegraphics[width=0.8\textwidth]{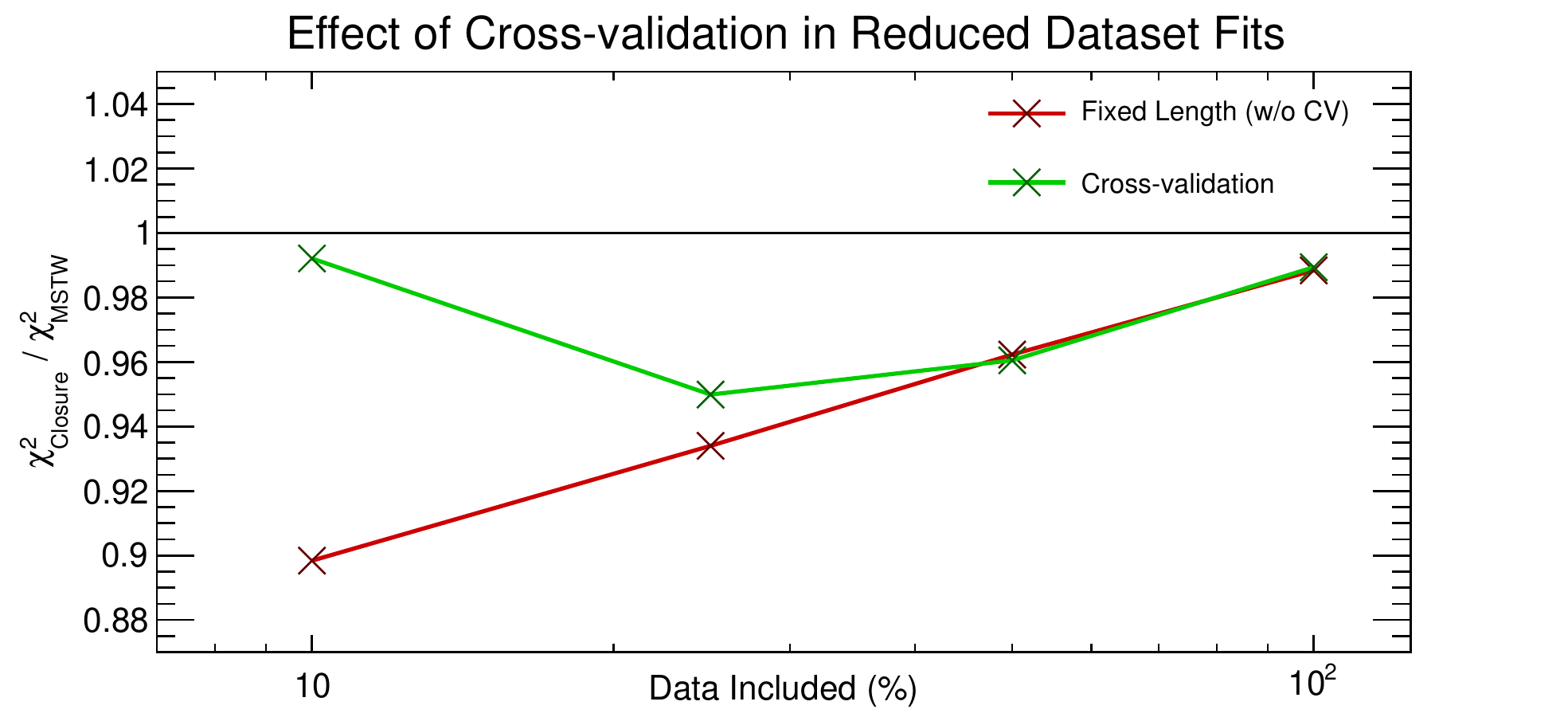}
\caption{\small Ratio of $\chi^2$s of closure test fits with and without cross-validation and the MSTW PDFs used to generate the closure test pseudo-data, for fits with reduced datasets.
Each fit has a dataset which has been randomly reduced to a specified percentage (100\%, 50\%, 25\% and 10\%).}
\label{fig:dchi2cv}
\end{figure} 

We can look at the impact of the new cross-validation on the NNPDF fits in closure tests.
Fig.~\ref{fig:diffdistance-cv-fl} shows the distances between the PDFs from a closure test fit using look-back cross-validation compared to a comparable fixed length fit. 
The distance is here defined as the absolute difference between the closure tests in units of the combined PDF uncertainty on each mean.
The closure test fits shown here were both performed using pseudo-data generated using the MSTW2008 PDFs. 
The distance between the two fits is generally below five for all PDFs, indicating that there is only a slight difference above what we would expect based on statistical fluctuations. 
%
These results tell us that that the introduction of look-back cross-validation has only a small impact on the full NNPDF3.0 fit.
This indicates that whatever overfitting is present in the NNPDF fits is small, possibly because of the large size of the dataset used, which has a high level of redundancy. 

However, there are still several reasons to include cross-validation in the final NNPDF3.0 settings.
It is possible that the tests we have used are not precise or comprehensive enough to detect all over-learning, and some could remain in the fit. 
Also, we would like to use the same methodology for all fits, including those with reduced datasets where over-learning may be a larger problem.
Fig.~\ref{fig:dchi2cv} demonstrates this by showing the $\chi^2$ ratio between the closure test and MSTW PDFs in fits where only a random subset of the dataset is used.
For large datasets (100\% and 50\%) there is little difference from including cross-validation, while in small datasets without cross-validation the $\chi^2$ is significantly smaller than the MSTW ideal, indicating over-learning.

\section{Closure testing LHC observables}

\begin{figure}
\centering
\includegraphics[width=0.45\textwidth]{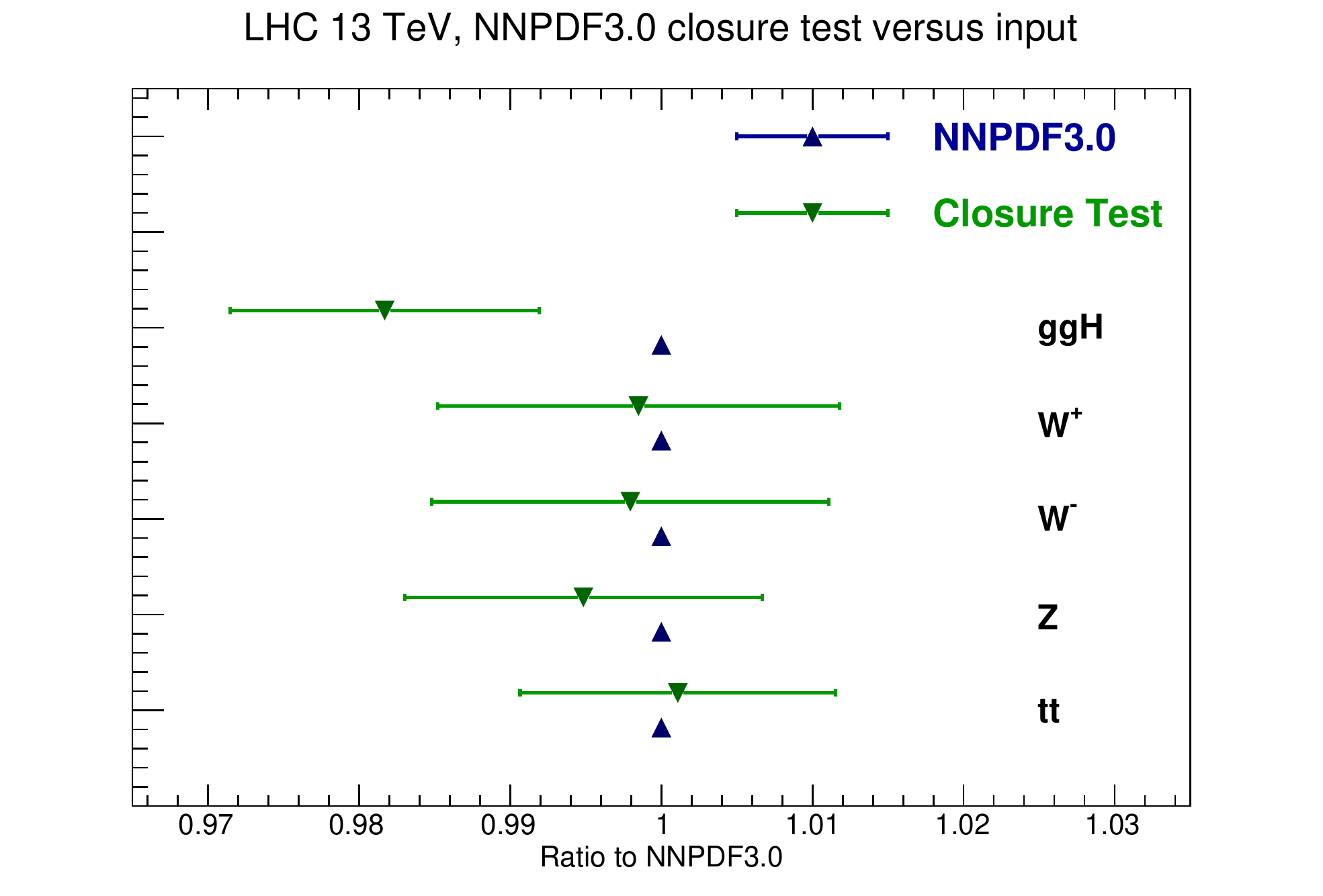}
\includegraphics[width=0.45\textwidth]{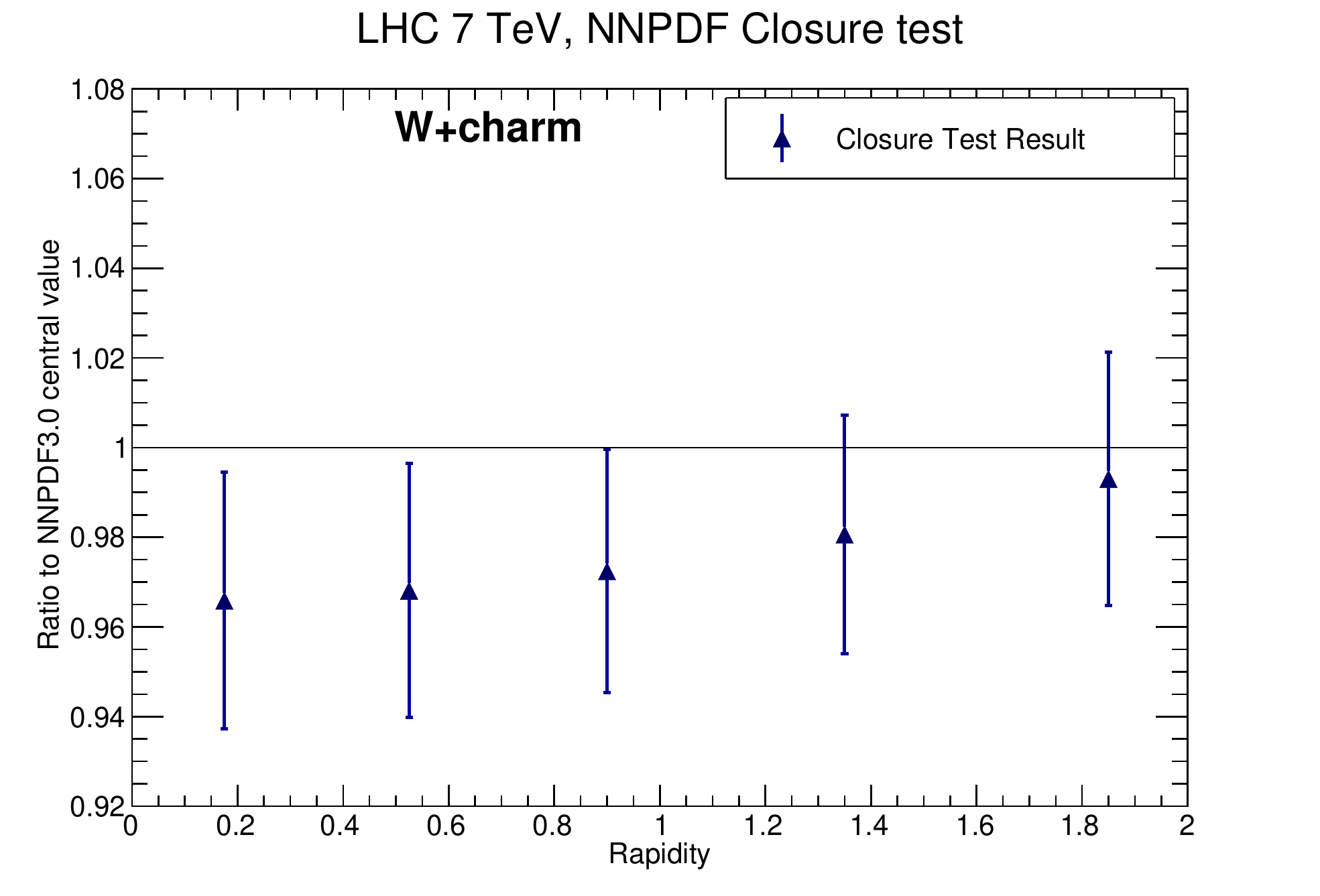}
\caption{\small (left) Comparison of closure test based on NNPDF3.0 pseudo-data to NNPDF3.0 for various 13 TeV LHC processes. (right) Same but for $W+c$ at different rapidity values.}
\label{fig:ctnnpdf}
\end{figure} 

The closure test results presented in~\cite{NN30} generally looked at the reproduction of the underlying PDFs at the level of the $\chi^2$ to the included experimental data and of the PDFs themselves. 
However, we can also look at how well LHC observables are reproduced in closure tests. 
Fig.~\ref{fig:ctnnpdf} shows calculations of a variety of LHC observables using a closure test fit based on NNPDF3.0-derived pseudo-data, compared to similar values calculated with the NNPDF3.0 PDFs themselves. 
The left-hand plot compares inclusive cross-sections for vector-boson production (computed with {\small\sc Vrap}~\cite{vrap}), top pair production ({\small\sc top++}~\cite{top++}), and Higgs production by gluon-gluon fusion ({\small\sc ggHiggs}~\cite{ggHiggs}), while the right-hand plot shows the differential cross-section for $W^++\bar{c}$ production. 
In most cases the closure test result is consistent with the input PDF value at the one-sigma level.
The largest difference is seen for the $ggH$ cross-section, where the closure test is about two standard deviations from the NNPDF3.0 value.
This level of difference is unsurprising given the statistical fluctuations in the closure test, as we expect, and have explicitly tested for the PDFs, that the one-sigma band contains the theory value in 68\% of cases.

\begin{figure}
\centering
\includegraphics[width=0.45\textwidth]{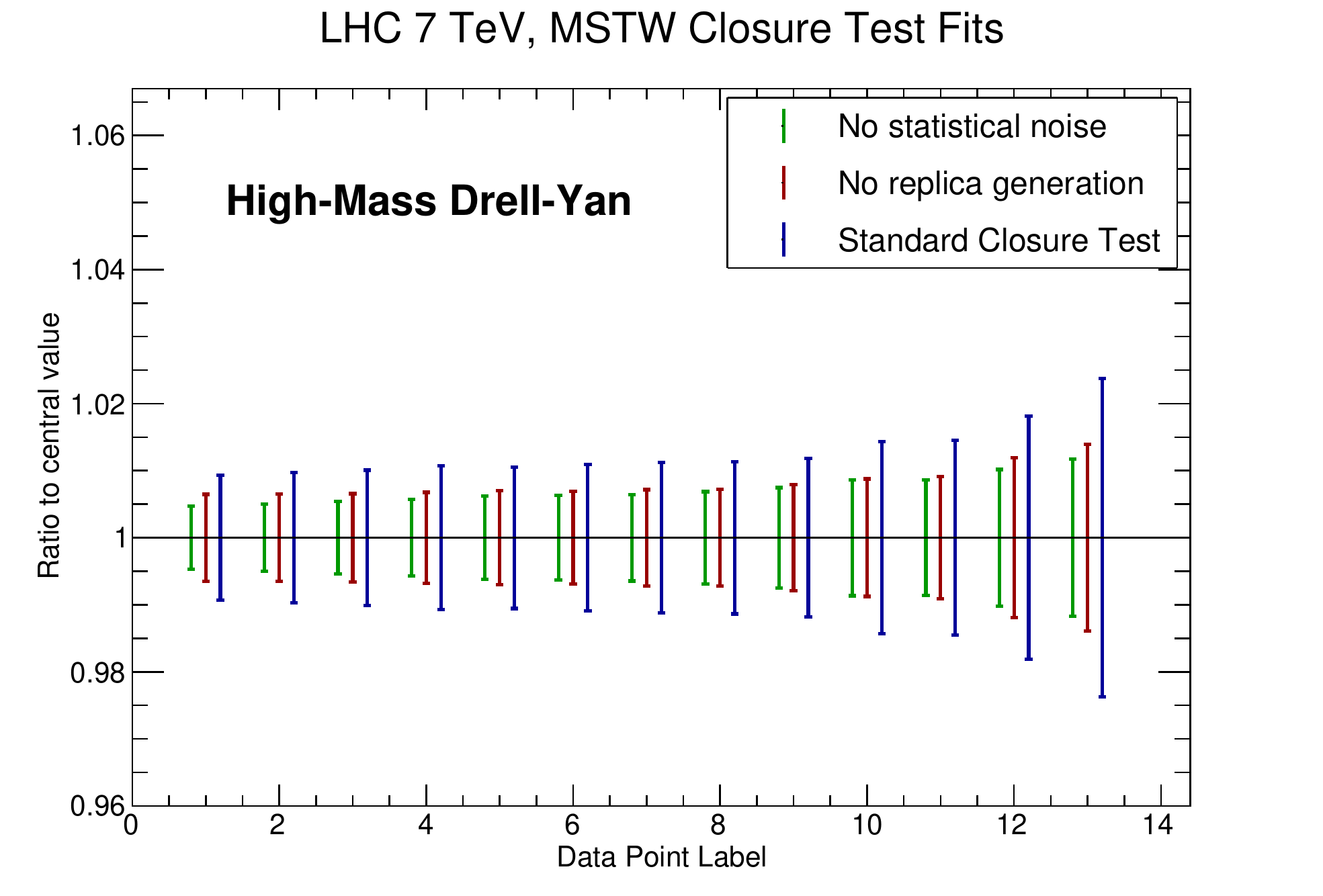}
\includegraphics[width=0.45\textwidth]{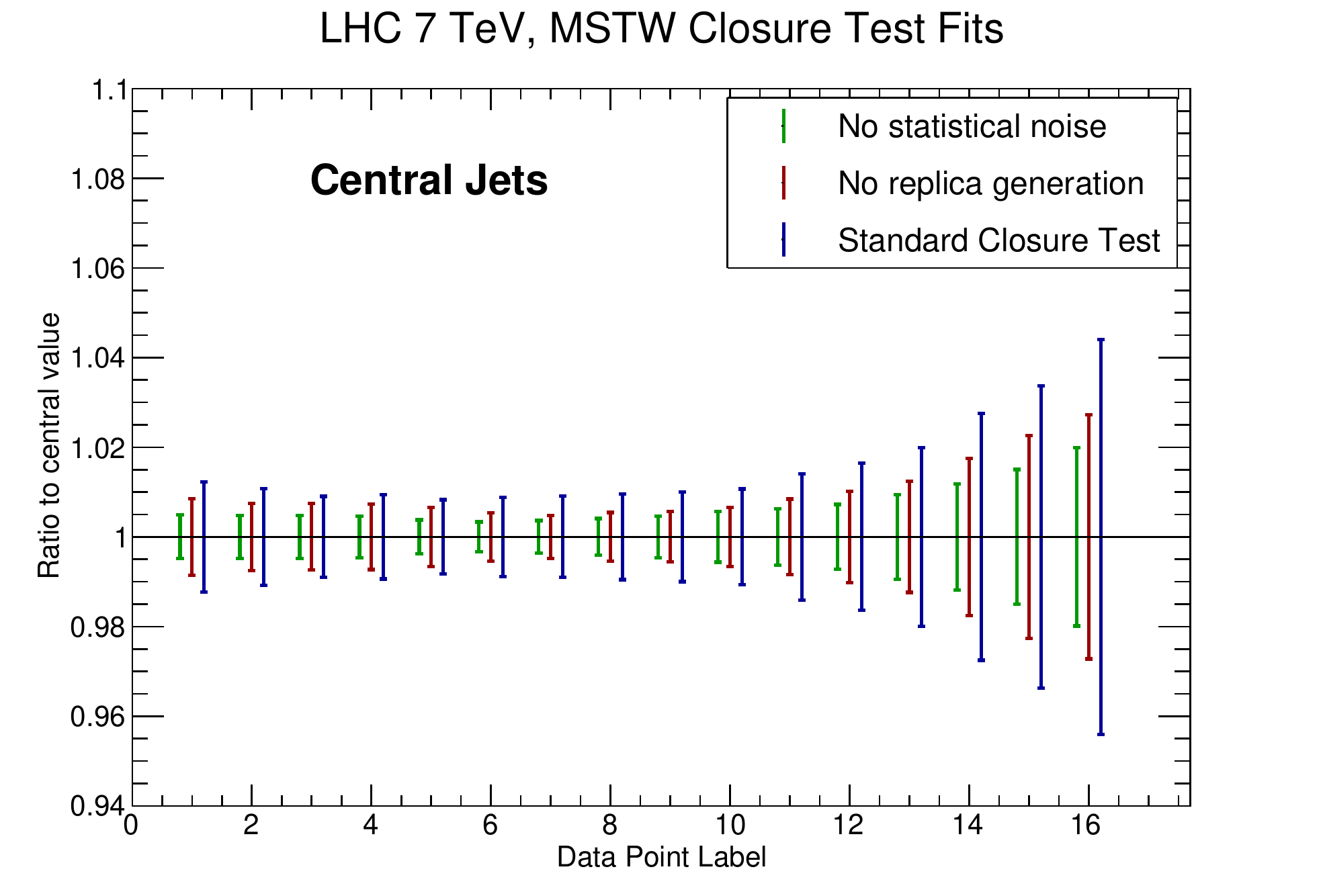}
\caption{\small (left) Comparison of the uncertainties in closure test fits performed with different levels of statistical noise for the data-points of the 7 TeV ATLAS high-mass Drell-Yan dataset. The uncertainties are shown for each fit as a ratio to the central value of the fit. (right) Same, but for central points in the ATLAS 7 TeV inclusive jet dataset.}
\label{fig:ctlevels}
\end{figure} 

We can use LHC observables to look at other closure test results.
Fig.~\ref{fig:ctlevels} compares the uncertainties obtained for ATLAS high-mass Drell-Yan and central inclusive jets in closure test fits performed with different levels of statistical noise.
This allows the different contributions to the PDF uncertainty to be disentangled; for instance the uncertainty obtained in the fit without any statistical noise can be identified as the extrapolation uncertainty, due to the limited resolution of the data.
Differences between successive levels of noise then provide an estimate of the functional uncertainty, from the existence different equally-probable PDFs, and finally data uncertainty, the propagation of the data uncertainties to the PDFs.
The results in Fig.~\ref{fig:ctlevels} are qualitatively similar to those found in~\cite{NN30} looking at the PDFs themselves, with the data uncertainty generally dominant, though with sizeable contributions from functional and extrapolation sources.
A more precise description of the different sources of uncertainties is given in the NNPDF3.0 paper~\cite{NN30}.

%

\end{document}